\documentclass[3p,times,procedia]{elsarticle}
\usepackage{nupha_ecrc}

\usepackage{amsmath}
\usepackage{graphicx}
\usepackage{subfig}

\volume{00}

\firstpage{1}

\journalname{Nuclear Physics A}

\runauth{}

\jid{nupha}

\jnltitlelogo{Nuclear Physics A}

\usepackage{amssymb}

\usepackage[figuresright]{rotating}

\begin{document}

\begin{frontmatter}



\dochead{XXVIIth International Conference on Ultrarelativistic Nucleus-Nucleus Collisions\\ (Quark Matter 2018)}

\title{Direct flow of heavy mesons as unique probe of the initial Electro-Magnetic fields in Ultra-Relativistic Heavy Ion collisions}


\author[add1,add2]{G. Coci} 
\author[add2,add3]{L. Oliva}
\author[add1]{S. Plumari}
\author[add1,add4]{S. K. Das}
\author[add1,add2]{and V. Greco}
\address[add1]{Department of Physics and Astronomy, University of Catania, Via S. Sofia 64, I-95125 Catania, IT}
\address[add2]{INFN-LNS, Laboratori Nazionali del Sud, Via S. Sofia 62, I-95125 Catania, IT}
\address[add3]{GSI Helmholtzzentrum f\"ur Schwerionenforschung GmbH, Planckstrasse 1, D-64291 Darmstadt, DE}
\address[add4]{School of Nuclear Science and Technology, Lanzhou University, 222 South Tianshui Rd, Lanzhou 730000, CN}

\begin{abstract}
In Ultra-relativistic Heavy-Ion Collisions (HICs) very strong initial electro-magnetic (E.M.) fields are created: the order of magnitude of the magnetic field is about $10^{19} \, Gauss$, the most intense field in the Universe, even larger than that of a magnetar. These fields rapidly decrease in time, inducing a drift of particles in the reaction plane. The resulting flow is odd under charge exchange and this allows to distinguish it from the large vorticity of the bulk matter due to the initial angular momentum conservation. Conjointly charm quarks, thanks to their large mass $M_{c}>>\Lambda_{QCD}$, are produced in hard partonic processes at formation time $\tau_f \approx 1\,/\,( 2M_{HQ} )$ which is comparable with the time scale when the E.M. field attains its maximum value. Moreover, with a mass of $M_c \approx 1.3 \,$ GeV there should be no mixing with the chiral magnetic dynamics and the condition $M_c \gg T$ allows charm quarks to have sufficiently large thermalization time, so that they can probe the entire phase-space evolution of the QGP retaining the initial kick given by the E.M. field. We show that such E.M. field entails a transverse motion of charm quarks resulting in a splitting of directed flow $v_1$ of $ D$ and $\bar{D}$ mesons of few percent, i.e. much larger compared to the measured pion one.
\end{abstract}


\end{frontmatter}


%


\section{E.M. fields in HICs: a realistic model}
\label{section1} 
\begin{figure}[h]
\centering
\includegraphics[width=0.45\textwidth]{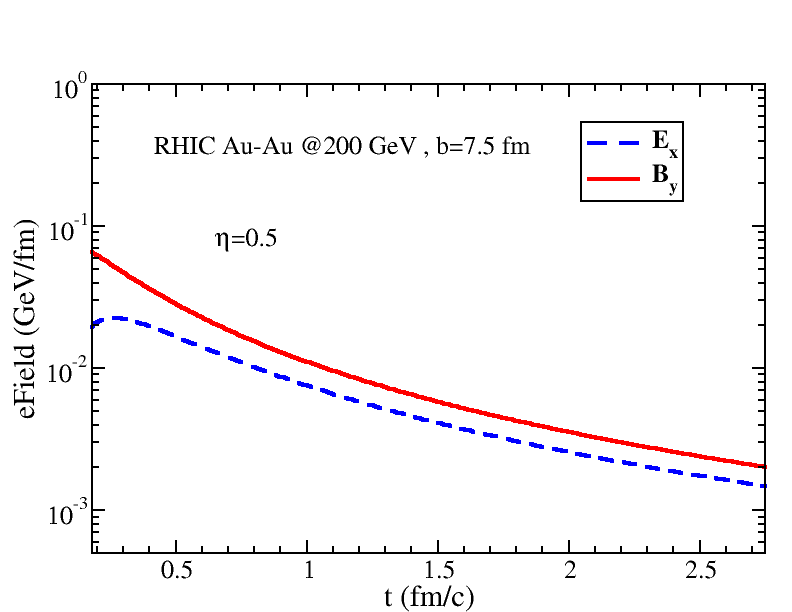}
\caption{Time profile of E.M. field main components $eE_x$ and $eB_y$ calculated for RHIC collisions at $b=7.5 \, fm$ for rapidity value $\eta=0.5$.}
\label{EMfieldRHIC}
\end{figure}

In the standard picture of HICs the colliding nuclei are composed objects made by point-like charges, the protons, which propagate along the \emph{z} direction at almost the speed of light. Choosing by convention the impact parameter \emph{b} along the \emph{x}-axis, the generated magnetic field $\vec{B}$ is dominated by the \emph{y}-component with an estimated initial value of $eB_y \approx 5 m_{\pi}^2$ and $eB_y \approx 50 m_{\pi}^2$ at RHIC and LHC energies respectively.
Time variation of $\vec{B}$ induces an electric field $\vec{E}$, whose dominant component is $E_x$, and results in a Faraday current $\vec{J}_{Faraday}=\sigma_{el} \vec{E}$ which drifts charged particles in the \emph{xz} plane. Here $\sigma_{el}$ is the electric conductivity of the QGP. Meanwhile, the Lorentz force $q\vec{v}\times\vec{B}$ acts on the expanding medium along the direction orthogonal to $\vec{B}$ and to the flow velocity $\vec{v}$ akin to the classical Hall current $\vec{J}_{Hall}$. The net combination of the two effects leads to the formation of a finite direct flow $v_1=<p_x/p_T>$~\cite{QGPCataniaPLB768}. 
\noindent Following Ref.~\cite{Gursoy2014} we derive the magnetic field from spectator protons $\vec{B}_s$ using the following formula
\begin{equation}\label{EMfield}
e\vec{B}_{s}=-Ze \int d\phi' dx'_{\perp} x'_{\perp} \rho_{-}(x'_{\perp},\phi') \left[\vec{B}_s^{+}(\tau,\eta,x_{\perp},\phi) + \vec{B_s}^{+}(\tau,-\eta,x_{\perp},\phi) \right]
\end{equation}
and an analogous one for the electric field $e\vec{E}_s$. In Eq.~\eqref{EMfield} $B_{s}^{+}$($B_{s}^{-}$) is the magnetic field generated by a single charge \emph{e} located at position $\vec{x}_{\perp}=(x_{\perp},\phi)$ in transverse plane and moving towards the \emph{+z}(\emph{-z}) direction with speed $\beta$ (rapidity $\eta=\arctan(\beta)$). These elementary E.M. fields are analytically calculated by solving Maxwell equations~\cite{Tuchin2013}, then folded with the nuclear transverse density $\rho_{-}(x_{\perp},\phi)$ and summed over forward ($\eta$) and backward ($-\eta$) rapidity. Participant protons lose some rapidity during collisions, so their contribution to the E.M. field is smoothed out and become secondary compared to the spectator part at least in the initial stage.
We assume constant $\sigma_{el}=0.023 fm^{-1}$, as predicted by lattice QCD (lQCD) calculations~\cite{Amato2013} around $T\approx 2T_c$. That means that we are actually neglecting any bulk modification due to E.M. currents. In this work we do not consider also fluctuations in event-by-event collisions that could cause other components of the E.M. field to become comparable with the dominant ones $B_y$ and $E_x$. An example of time evolution of the obtained E.M. field obtained is shown in Fig.~\eqref{EMfieldRHIC}.
\vspace*{-0.4cm}
\vspace*{-0.2cm}
%
%

\section{Boltzmann approach for dynamical evolution of HQs in QGP}
\label{section2}
We describe the propagation of HQs in the QGP by means of the relativistic Boltzmann equation
\begin{equation}\label{BoltzmannEM}
\left[ p_{\mu}\partial_{x}^{\mu} + q F_{\mu\nu}(x)p^{\nu}\partial_{p}^{\mu}\right]f_{HQ}(x,p) = C_{22}[f_{HQ}(x,p),f_{g}(x,p),f_{q}(x,p)]
\end{equation}
In order to solve numerically Eq.~\eqref{BoltzmannEM} we divide the coordinate space in a 3D grid and we sample the single particle phase-space distribution function $f(x,p)$ using test-particle method.
On the left-hand side of Eq.~\eqref{BoltzmannEM} the Maxwell strength tensor $F_{\mu\nu}$ is constructed using $\vec{E}$ and $\vec{B}$ from previous section. On the right-hand side the Boltzmann-like collision integral $C_{22}[f_{HQ},f_{g},f_{q}]$ encodes the dissipative interactions between HQs and bulk partons and it is mapped into a collision probability by means of a stochastic algorithm. In this work we consider only elastic processes using scattering matrices calculated at Leading-Order of pQCD. Moreover, we take into account non-perturbative effects by means of Quasi-Particle (QP) prescription~\cite{QGPCataniaPRD84}: bulk partons are dressed with thermal masses $m_{g,q}(T)\propto g(T)\,T$
and the \emph{T}-dependence of the coupling $g(T)$ is tuned to lQCD thermodynamics~\cite{Borsanyi2010}. 
In realistic simulations at RHIC  we distribute charm quarks in momentum space according to Fixed Order + Next-to-Leading-Order (FONLL) \emph{pp}-spectra \cite{FONLL RHIC (2005)}. 
For bulk partons we employ Boltzmann-J\"uttner distribution plus minijet tail at high $p_T$. In coordinate space we provide initial conditions through standard Glauber model with a slight modification.
In order to account the partial transfer of the angular momentum of the two nuclei to the plasma, we modify our usual equilibrium initial condition adding a velocity profile which varies along the direction of the impact parameter $b$. We choose such velocity profile in agreement with other models \cite{vorticity4} \cite{vorticity5} \cite{vorticity6} for what concern the local vorticity distribution in the fireball and the total angular momentum induced by the two colliding nuclei to the system and retained by the plasma as a shear flow in the longitudinal direction. 
As last ingredient, at the final stage of the evolution we couple an hadronization mechanism for HQs which is based on a hybrid fragmentation plus coalescence model \cite{QGPCataniaEPJC78}.
Within our Boltzmann approach we are able to simultaneously describe the nuclear suppression factor $R_{AA}(p_T)$ and the elliptic flow $v_2(p_T)$ of \emph{D} mesons both at RHIC and LHC energies \cite{QGPCataniaPRC96}. Here we point out that the inclusion of E.M. field does not produce a significant effect neither on  $R_{AA}$ nor on $v_{2}$. Hence, we guarantee to focus only on the direct flow as a promising observable for probing the initial E.M. fields~\cite{QGPCataniaPLB768}. 
\section{Results and Conclusions}
\label{section3}
\begin{figure}[h]
\centering
\subfloat[][]{\includegraphics[width=.45\textwidth]{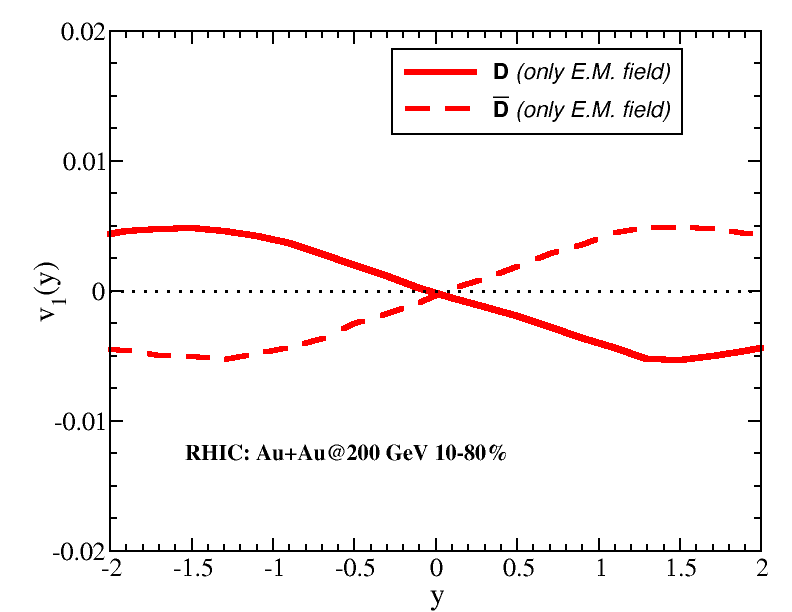}}\quad
\subfloat[][]{\includegraphics[width=.45\textwidth]{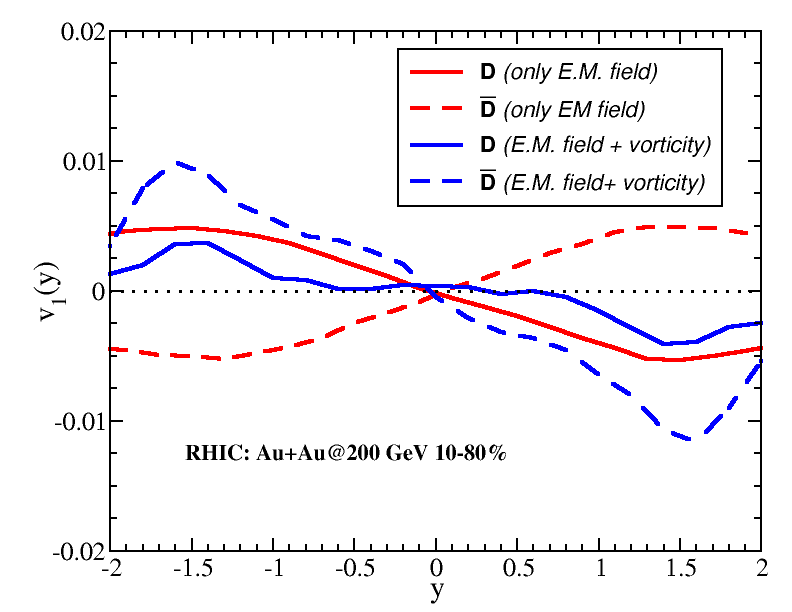}} 
\caption{Predicted direct flow $v_1$ of  $D$-$\bar{D}$ mesons as function of rapidity \emph{y} at RHIC collisions with $b=9 \, fm$: (a) $v_1$ produced due to initial E.M. field, (b) $v_1$ obtained in the E.M. field plus vorticity coupling configuration compared to only E.M. field scenario.} 
\label{v1RHIC} 
\end{figure}

In Fig.~\eqref{v1RHIC} we present our predictions for direct flow $v_1$ of \emph{D} mesons at RHIC events. The effect of the initial E.M. field results into a $v_1 \approx 10^{-2}$, i.e. much larger than  $v_1$ in the light sector. It is also visible the rapidity and charge odd-dependence of $v_1$, which seems to be maintained also in a scenario where we couple the vorticity distribution.  
In summary, we studied the dynamics of HQs within a Boltzmann approach where we account non-perturbative interaction through a QP prescription tuned to lQCD equation of state.    
In accordance to the hint of Ref.~\cite{QGPCataniaPLB768} this work suggests that the effect of a strong E.M. field created at initial stage of HICs shows off directly the quark degrees of freedom of the QGP. 
In particular, the $v_1$ of HQs appears to be an efficient probe as the observation of splitting of uncharged $D^0$-$\bar{D}^0$ would be a clear signature of the deconfined phase.
In future we will focus on the possibility to implement a more complex vorticity model~\cite{Chatterjee2018}, to include \emph{T}-dependent electric conductivity and add event-by-event fluctuations.      

\vspace*{-0.2cm}




\bibliographystyle{elsarticle-num}



\end{document}